\documentstyle[12pt]{article}
\newcommand{\uq}{U_q(g)}
\newcommand{\ur}{U_q^{(r)}(gl(2|1))}

\newcommand{\lam}{\lambda}
\newcommand{\Lam}{\Lambda}
\newcommand{\x}{\otimes}

\newcommand{\ba}{\begin{eqnarray}}
\newcommand{\na}{\end{eqnarray}}
\newcommand{\ban}{\begin{eqnarray*}}
\newcommand{\nan}{\end{eqnarray*}}

\newcommand{\Z}{{\bf Z}_2}
\newcommand{\Kr}{K_{2\rho}}
\newcommand{\Rep}{Rep_A}
\newtheorem{lemma}{Lemma}
\newtheorem{theorem}{Theorem}

\begin{document}
\title{ Quantum Supergroups And\\
Topological Invariants of Three - Manifolds}
\author{R. B. Zhang\\
\small  Department of Pure Mathematics\\
\small University of Adelaide\\
\small Adelaide, SA 5001, Australia}
\maketitle

\vspace{3cm}
\noindent
The Reshetikhin - Turaeve approach to topological invariants
of three - manifolds is generalized to quantum supergroups.
A general method for constructing  three - manifold invariants
is developed, which requires only the study of the eigenvalues
of certain central elements of the quantum supergroup in
irreducible representations. To illustrate how the method works,
$U_q(gl(2|1))$ at odd roots of unity is
studied in detail, and the corresponding topological invariants
are obtained. \\

\pagebreak
\section{Introduction}
The discovery of the Jones invariants\cite{Jones}
marked the beginning of a new phase in knot theory.
Soon after Jones discovery, it was realized\cite{Kauffman}
that these invariants are intimately connected with soluble models
in statistical mechanics\cite{Baxter}
through the Yang - Baxter equation\cite{Yang}\cite{Baxter}.
Subsequent investigations on this
connection\cite{Turaev}\cite{Wadati}  turned out to be very fruitful.
The connection of knot theory and physics went much further.
In a seminal paper, Witten interpreted  the Jones and related
link polynomials in terms of the quantum Chern - Simons theory\cite{Witten}.
The virtue of Witten's formulation is that it is intrinsically three
dimensional,  thus topological invariants of
three manifolds are obtained in the same token.

In the study of knot theory and the Yang - Baxter equation,
quantum groups\cite{Drinfeld}\cite{Jimbo} and quantum
supergroups\cite{supergroups} played an important role.
A quantum (super)group $\uq$ has the properties of a ($\Z$ graded)
quasi triangular Hopf algebra, admitting a universal
$R$ - matrix, which satisfies the Yang - Baxter equation.
Categorical arguments show\cite{ReT} that every tangle can be
associated with a homomorphism of finite dimensional
representations of $\uq$, in a manner consistent with
the isotopy properties of the former.
In particular, a link is associated with a
map from the trivial $\uq$ module to itself, which yields a regular
isotopy invariant of the link(a simple scaling of the map produces
an ambient isotopy invariant)\cite{Res}.
A more down to earth approach\cite{ZGB} to link polynomials,
which is closer to Jones'
original construction in spirit,  uses the fact that in any
finite dimensional irreducible representation of $\uq$,
the $R$ - matrix generates a representation of the braid group,
on which a Markov trace can be defined in a general fashion.
In this way, each finite dimensional irreducible representation of
any quantum (super)group leads to a link invariant, and all
the well known link polynomials can be obtained as special cases.
For example, the Homfly polynomial\cite{Homfly} arises
from the vector representations of $U_q(gl(m))$ and $U_q(gl(m|n))$,
$m\ne n$(with the Jones polynomial as a special case),
the Alexander - Conway polynomial from the vector representation of
$U_q(sl(m|m))$, and the Kauffman polynomial from
the vector representations of $U_q(so(n))$ and $U_q(osp(M|2n))$.

It is well known that the study three -  manifold topology
may be reduced to the study of links embedded in $S^3$.
In the early 1960s, Lickorish and Wallace established a theorem
stating that each framed link in $S^3$ determines a compact, closed,
oriented $3$ - manifold, and every such $3$ - manifold is obtainable
by surgery along a framed link in $S^3$\cite{Lic}.
The problem with this description of $3$ -  manifolds is that it is
not unique: different framed links may yield homeomorphic $3$ - manifolds
upon surgery.  This problem was resolved by Kirby\cite{Kirby} and Craggs,
whose result was further refined by Fenn and Rourke in \cite{FR}.
These authors proved that orientation preserving homeomorphism classes
of  compact, closed, oriented $3$ - manifolds correspond bijectively
to equivalence classes of framed links in $S^3$, where the equivalence
relation is generated by the Kirby moves.

In view of these results, one naturally expects it to be possible to
construct topological invariants of $3$ - manifolds using the link
polynomials arising from quantum (super)groups as building blocks,
and this is indeed the case.
In a seminal paper\cite{RT},  Reshetikhin and Turaev
developed a purely algebraic approach to three - manifold
invariants based on the theory of quantum groups.
The essential  idea of \cite{RT} is to make appropriate
combinations of isotopy invariants of a framed link embedded in $S^3$,
such that they will be intact under the Kirby moves, thus qualify
as topological invariants of the three - manifold obtained
by surgery along this link. In \cite{RT}, a method was also deviced
for explicitly constructing such combinations. Applying it to
the quantum group $U_q(sl(2))$ at even  roots of unity
led to a set of three - manifold invariants, which turned out to
be equivalent to those obtained by Witten using $sl(2)$ Chern -
Simons theory\cite{Singer} -- \cite{Rozansky}.
The application of this method in general requires
detailed analysis of indecomposable representations of, typically,
a quantum group at roots of unity, a problem of a very high degree
of difficulty. However, guided by properties of Witten's Chern -
Simons theory, Turaev and Wenzl in a remarkable paper\cite{TW}
resolved the problem for $U_q(gl(m))$ and $U_q(so(n))$ at roots
of unity, thus proved the existence  of the three - manifold
invariants associated to these quantum groups.

The algebraic  approach of Reshetikhin and Turaev was generalized
to the quantum supergroup $U_q(osp(1|2))$ at odd roots of unity in
\cite{inv}, and the corresponding three - manifold invariants
were constructed. In that paper, a way was found to avoid
the difficult problem of  classifying indecomposable
representations. The construction there required only the
study of the eigenvalues of certain central elements of the
algebra in the irreducible representations.

We should  point out that the Witten - Reshetikhin - Turaev invariant was
also provided with an elementary derivation by Lickorish\cite{Lic1},
who utilized only properties of the Temperley - Lieb algebra.
We should also mention that there exists another
well known $3$ - manifold invariant due to Tureav and Viro\cite{Viro},
which was obtained from the quantum group analog of Ponzano and Regge's
formula for the partition function of three - dimensional quantum
gravity on a lattice\cite{Regge}.
The various approaches to $3$ -  manifold invariants
are undoubtedly related in an intimate way, and
a coherent understanding of their connection will be very interesting.

The aim of this paper is to generalize the Reshetikhin - Turaev approach
to ${\bf Z}_2$ graded Hopf algebras, thus to erect a general framework for
constructing three - manifold invariants using quantum supergroups.
As is well known, the representation theory of quantum supergroups
is drastically different from that of their nongraded counter parts,
hence we expect that quantum supergroups will yield
three - manifold invariants genuinely different from those
derived from ordinary quantum groups.
Another fact worth noticing is that there exist severe difficulties in
developing a proper formulation of Chern - Simons theory with super
gauge groups\cite{SCS}. Therefore the algebraic approach seems to be
a more practicable way to the construction of three - manifold invariants
using supergroups.

A method for constructing  Kirby move invariant
combinations of link polynomials will be given in section 4,
which, as we will see, essentially boils down to finding a set of
constants $d_\lam$'s, which render vanishing the central element
$\delta$ (see equation (\ref{eq:delta}) ) of the employed
${\bf Z}_2$ graded Hopf algebra $A$ in all irreducible $A$ modules
with  nonzero $q$ - superdimensions.
This construction is applied to the quantum supergroup
$U_q(gl(2|1))$  at odd roots of unity in section 5,
and the corresponding three - manifold invariants are obtained.
To make our paper self contained, we review
the basic  properties of coloured ribbon graphs\cite{ReT},
and also present some notions of ${\bf Z}_2$ - graded Hopf
algebras(i.e., Hopf superalgebras).

\section{$\Z$ graded Hopf algebras}
Let us begin by quickly reviewing the definition of a $\Z$ graded
Hopf algebra.  We will work in the complex field ${\bf C}$
throughout the paper.
A $\Z$ graded vector space over $\bf C$ is a dirct sum of two
subspaces $V=V_{0}\oplus V_{1}$, with $V_{0}$ and $V_{1}$ called
the even and odd subspaces respectively. We introduce the gradation
index $[]: V_{0}\cup V_{1}\rightarrow \Z=\{0,1\}$, such that if
$x\in V_{i}$, then $[x]=i$. We will call an element of $V$
homogeneous if it belongs to $V_{0}\cup V_{1}$, and inhomogeneous
otherwise. The dual space $V^*$ of $V$ is also $\Z$ graded,
$V^*=V_{0}^*\oplus V_{1}^*$, where $V_i^*=Hom_{\bf C}(V_i, \bf C)$,
$i\in\Z$. The tensor product $V\x_{\bf C} W$ of two $\Z$ graded vector
 spaces $V$ and $W$ inherits a $\Z$ grading from that of $V$ and $W$,
with $V\x W=\oplus _{i\in\Z}(V\x W)_i$, $(V\x W)_i=\oplus_{k}
V_k\x W_{i-k(mod 2)}$.

A linear map $f: V\rightarrow W$ of two $\Z$ graded vector
spaces $V$ and $W$ is said to be homogeneous of degree $r\in\Z$ if
$f(V_i)\subseteq W_{i+r(mod 2)}$. We will call $f$ a homomorphism if
$r=0$, an isomorphism if it is also one - to - one and onto.
Note that the linear homogeneous map  $f: V\rightarrow W$
is uniquely defined by specifying the images of
the homogeneous elements of $V$, then extending the definition to the
entire space through linearity.
For later use, we now define the twisting  homomorphism
\ba
T: V\x W\rightarrow W\x V.
\na
All we need to do is to designate for any homogeneous elements
$x\in V$ and $y\in W$ that
\ban
T(x\x y)=(-1)^{[x][y]}y\x x,
\nan
and then extends $T$ to all elements of $V\x W$ through linearity.

A $\Z$ graded algebra $A$ is a $\Z$ graded vector space equipped with
homomorphisms $M: A\x A\rightarrow A$, and $e: \bf C\rightarrow A$,
respectively called the multiplication and unit, such that $M$
is associative, i.e., $M(M\x id)=M(id\x M)$, and $e(c)a=a e(c)=c a$,
where $c\in {\bf C}$, $a\in A$, and $id: A \rightarrow A$ is the
identity map. Note that the tensor
product $A\x_{\bf C} B$ of two $\Z$ graded algebras is again a $\Z$
graded algebra, with the multiplication given by
\ban
(a_1\x b_1).(a_2\x b_2)&=&(-1)^{[a_2][b_1]}a_1a_2\x b_1b_2,
\nan
where $a_1, a_2\in A$, $b_1, b_2\in B$.
Let $f: A\rightarrow B$
be a homomorphism of the underlying $\Z$ graded
vector spaces. We call $f$ a $\Z$ graded algebra homomorphism if
for $a_1, a_2\in A$,
\ban
f(a_1a_2) = f(a_1)f(a_2),
\nan
and an anti - homomorphism if
\ban
f(a_1a_2) = (-1)^{[a_1][a_2]}f(a_2)f(a_1).
\nan

A $\Z$ graded co - algebra $B$ is a $\Z$ graded vector space equipped
with homomorphisms $\Delta: B\rightarrow B\x_{\bf C} B$, and $\epsilon:
B\rightarrow \bf C$, respectively called the co - multiplication and
co - unit, such that $\Delta$ is co - associative, i.e.,
\ban
(\Delta\x id)\Delta = (id\x \Delta)\Delta,
\nan
and $\epsilon$ has the following unitarity property
\ban
(\epsilon\x id)\Delta = (id \x \epsilon)\Delta = id.
\nan
The tensor product
$B\x_{\bf C} A$ of
two $\Z$ graded co - algebras $B$ and $A$ is again a $\Z$ graded co -
algebra with the co - multiplication $(id_B\x T  \x id_A)(\Delta_B\x
\Delta_A)$ and co - unit $\epsilon_B\x\epsilon_A$.

Let $A$ be a $\Z$ graded algebra with multiplication $M$ and unit $e$,
and at the same time also be a $\Z$ graded co - algebra with co -
multiplication $\Delta$ and co - unit $\epsilon$. We call $A$ a
$\Z$ graded bi - algebra if both $\Delta$ and
$\epsilon$ are $\Z$ graded algebra homomorphisms.
If further, there exists an anti - homomorphism $S: A \rightarrow A$,
such that $M(id\x S)\Delta$  $= M(S\x id)\Delta$ $=\epsilon$,
then $A$ is called a $\Z$ graded Hopf algebra, or a Hopf superalgebra.

Let us introduce the category of finite
dimensional linear representations $\Rep$ of $A$, of which the
objects are the finite dimensional left $A$ modules over $\bf C$,
and the morphisms are $A$ linear homomorphisms.
Note that a given object $V$ in $\Rep$ is a $\Z$ graded vector space, i.e.,
$V=V_{0}\oplus V_{1}$. Let $a$ be a homogeneous element of $A$ with
$[a]=i$, then
\ban a: V_{j}\rightarrow V_{i+j(mod 2)}. \nan

Given two objects $V$ and $W$ of $\Rep$, the $\Z$ graded vector space
$V\x_{\bf C}W$
also belongs to $\Rep$, with the action of $a\in A$ defined by
\ban
a(x\x y) &=&\sum_{(a)}(-1)^{[a_{(2)}][x]}a_{(1)}x
\x a_{(2)}y,\ \ \ x\in V, y\in W,
\nan
where Sweedler's sigma notation
\ban \Delta(a)&=&\sum_{(a)}a_{(1)}\x a_{(2)},  \ \ \ a\in A,
\nan
has been employed.  For any $A$ module $V$, the dual vector space
$V^*$ is also an $A$ module with the action of $a\in A $ defined by
\ban
(a x^*)(y)&=&(-1)^{[a][x^*]} x^*(S(a)y), \ \ x\in V^*,  \ \ y\in V.
\nan

By the trivial $A$ module we shall mean the one dimensional module
${\bf C} x$ such that $a x=\epsilon(a) x$, $\forall a\in A$.

A quasi triangular $\Z$ graded Hopf algebra $(A, \ R)$ is a
Hopf superalgebra $A$, which admits an even element $R\in A\x A$,
called the universal
$R$ matrix,  satisfying the following relations
\ba
R\Delta(a)&=&\Delta'(a)R, \ \ \forall a\in A, \\
(\Delta\otimes id)R&=&R_{13}R_{23},\nonumber\\
(id\otimes\Delta)R&=&R_{13}R_{12},
\na
where $\Delta'=T\Delta$.
It is not difficult to show that the above equations imply
that $R$ obeys the Yang--Baxter equation
\ba
R_{12}R_{13}R_{23}=R_{23}R_{13}R_{12},
\na
and also satisfies
\ban
R^{-1}&=(S\x id)R&=(id\x S^{-1})R.
\nan
Express the universal $R$ matrix as $R=\sum_t\alpha_t\x\beta_t$,
and define
\ban
u&=&\sum_t (-1)^{[\alpha_t]}S(\beta_t)\alpha_t.
\nan
Then
\ba
u^{-1}&=&\sum_t(-1)^{[\alpha_t]}\beta_t S^2(\alpha_t),\nonumber \\
S^2(a)&=&u a u^{-1},   \ \ \ \forall a\in A.
\na

A ribbon Hopf superalgebra $(A, \ R, \ \Kr)$ is a quasi triangular
Hopf superalgebra $(A, \ R)$
with a group like element $\Kr$(i.e., invertible and
$\Delta(\Kr)=\Kr\x\Kr$),  such that
\ban
S^2(a)&=&\Kr a \Kr^{-1}, \ \ \ \forall a \in A.
\nan
Define
\ba
v&=&u\Kr^{-1}.\label{eq:v}
\na
It can be proved that $v$ lies in the central algebra of $A$, and
\ban
\Delta(v)&=&(v\x v) (R^TR)^{-1},
\nan
where $R^T=T(R)$.

Let $V$ be a finite dimensional $A$ module. We denote the
corresponding representation by $\pi$.
Define\cite{ZG}
\ba
C_V&=&Str_V[(\pi\x id)(\Kr^{-1}\x 1)R^TR],
\label{eq:C}
\na
where $Str_V$ represents the supertrace taken over $V$.
Then $C_V$ belongs to the central algebra of $A$.
The eigenvalue
of $C_V$ in the trivial irrep of $A$ is called the $q$
- superdimension of $V$, which we denote by $SD_q(V)$.

\section{Coloured ribbon graphs}
\subsection{Coloured ribbon graphs}
Coloured ribbon graphs were discussed extensively by Reshetikhin
and Turaev in \cite{ReT}and \cite{RT}. In order to make the present
paper self contained, we rephrase some of the basic notions here,
and refer to the above mentioned papers for further details.

By a ribbon we mean the square $[0,1]\times[0,1]$ smoothly embedded
in $R^3$. The images of $[0,1]\times 0$ and  $[0,1]\times 1$ are
the bases, and that of ${{1}\over {2}}\times [0,1]$ is called
the core of the ribbon.  Similarly an annulus is the cylinder
$S^1\times [0,1]$
embedded in $R^3$, and the image of $S^1\times  {{1}\over {2}}$
under the embedding is called the core of the annulus. Ribbons and
annuli are oriented as surfaces and their cores are directed.

Given $k, \ l\in {\bf Z}_+$.  A $(k, l)$ ribbon graph is an oriented
surface consisting of ribbons and annuli such that ribbons and annuli
never meet, and this surface intersects $({\bf R}^2 \times 0)\cup
({\bf R}^2 \times 1)$ in the bases of the ribbons, where the collection
of these bases are the collection of segments
$\{[i-{{1}\over{4}}, i+{{1}\over{4}}]\times 0\times 0 | i=1,2,...,k\}
\cup \{[j-{{1}\over{4}}, j+{{1}\over{4}}]\times 0\times 1 | j=1,2,...,l\}$.
For simplicity, we will represent a ribbon or an annulus by its
directed core.

We introduce two operations, composition and juxtaposition,
to manufacture new ribbon graphs from given ones.
Given $(k,\ l)$ graph $\Gamma_1$, $(l,\ m)$ graph $\Gamma_2$,
and $(k',\ l')$ graph  $\Gamma_3$,
the composition of  $\Gamma_1\circ\Gamma_2$ is defined
in the following way: shift  $\Gamma_2$ by the vector $(0,0,1)$
into ${\bf R^2}\times [1,2]$, glue the bottom end of
 $\Gamma_2$ to the top end of  $\Gamma_1$ in such a way that
the core of the ribbons glued together should have the same
direction as the core of the resultant ribbon(if
this is not possible, then the composition is not defined.),
then reduce the size of the resultant picture  by a factor of
$2$, leading to a $(k, \ m)$ graph.
The juxtaposition  $\Gamma_1 \otimes  \Gamma_3$ is to position
 $\Gamma_3$ on the right of  $\Gamma_1$, leading to a
$(k+k',\ l+l')$ graph.

By repeatedly applying these two operations to the
set of ribbon graphs depicted in Figure 1,
we can generate all the ribbon
graphs:

\vspace{1 cm}
\begin{center}
Figure 1.
\end{center}
\vspace{1 cm}

We associate to each $(k, \l)$ graph $\Gamma$   two sequences
$\epsilon_*(\Gamma) = (\epsilon_1, \epsilon_2, ..., \epsilon_k)$
and $\epsilon^*(\Gamma)=(\epsilon^1, \epsilon^2, ..., \epsilon^l)$,
$\epsilon^i, \epsilon_j \in \{1, \ -1\}$, in the following way.
For a ribbon of $\Gamma$ with a base $[i-{{1}\over{4}},
i+{{1}\over{4}}]\times 0\times 0$
(resp.  $[j-{{1}\over{4}}, j+{{1}\over{4}}]\times 0\times 1$),
if its core is directed towards(resp. away from) this base,
then $\epsilon_i=1$ (resp. $\epsilon^j=1)$, and $\epsilon_i=-1$
(resp. $\epsilon^j=-1)$ otherwise.

Let $(A,\ R, \ \Kr)$ be a ribbon Hopf superalgebra.
Let $\{V_i| i\in I\}$ be a set of finite dimensional $A$ modules.
We also introduce the set $\cal N$ consists of finite sequences
of the form  $((i_1, \ \epsilon_1 ), (i_2, \ \epsilon_2 ),...,
(i_k, \ \epsilon_k))$, $k\in {\bf Z}_+$, $\epsilon_t\in \{1, \ -1\}$.

A colouring of a ribbon graph $\Gamma$ is a mapping $col$
associating with each ribbon or annulus of $\Gamma$ an index $i\in I$.
The category $\cal H$ of coloured ribbon graphs is defined to have
as objects the elements of $\cal N$,
and as morphisms the coloured ribbon graphs, where we require
that if the coloured ribbon graph $\Gamma$ is a morphism
$\eta \rightarrow \eta'$, $\eta, \eta'\in {\cal N}$,  then the
sequences of colours and directions of cores of the bottom and top
ribbons must be equal to $\eta$ and $\eta'$ respectively.
$\cal H$ is provided
with a tensor product structure $\bigotimes: {\cal H}\times{\cal H}
\rightarrow {\cal H}$,  such that the tensor product of objects
$\eta$ and $\eta'$ is to position the latter on the right of $\eta$
to form one sequence, and the tensor product of morphism is simply
the juxtaposition of ribbon graphs defined earlier.

\subsection{The Reshetikhin - Turaev functor}
Reshetikhin and Turaev\cite{ReT} proved that there exists a unique covariant
functor from  $\cal H$ to the category of finite dimensional
representations of ribbon Hopf algebras, e.g., quantum groups.
Their result can be directly generalized to ribbon Hopf superalgebras,
leading to the following theorem
\begin{theorem}
Let $(A, \ R, \ \Kr)$ be a ribbon Hopf superalgebra,
$\{V_i | i\in I\}$ a set of finite dimensional $A$ modules,
and $\cal H$ the corresponding category of coloured ribbon
graphs. Then there exists a unique covariant functor $F: {\cal H}
\rightarrow \Rep$ such that
\begin{enumerate}
\item $F$ transform any object $\eta=
((i_1, \ \epsilon_1 ), (i_2, \ \epsilon_2 ),..., (i_k, \ \epsilon_k)) $
of $\cal H$ into the $A$
module $V(\eta)=V_{i_1}^{\epsilon_1}\otimes V_{i_2}^{\epsilon_2}\otimes
... V_{i_k}^{\epsilon_k}$, where $V^{+1}_i=V_i$,
 $V^{-1}_i=V^*_{i}$, and if $k=0$, then $V(\eta)$ is defined as the trivial
$A$ module.
\item For any two coloured ribbon graphs $\Gamma$ and $\Gamma'$,
\ba
F(\Gamma\otimes\Gamma')&=&F(\Gamma)\otimes F(\Gamma'),
\na
that is,   $F$ preserves the tensor product operation.
\item Colour the bottom left ribbons of $X^+$ and $X^-$ by
$i$ and the bottom right ones by $j$,  and denote the resultant
coloured ribbon graphs by $X^+_{ij}$ and $X^-_{ij}$ respectively,
then
\ba
F(X^+_{ij})=P R: & V_i\otimes V_j \rightarrow V_j\otimes V_i,\nonumber \\
F(X^-_{ij})=R^{-1}P: & V_i\otimes V_j \rightarrow V_j\otimes V_i,
\na
where $R$ is the universal $R$ matrix of $A$, and $P$ is the graded
permutation operator. Similarly, we have
\ba
F(I^+_i)=id:& V_i\rightarrow V_i,\nonumber\\
F(I^-_i)=id:& V^*_i\rightarrow V^*_i,
\na
and
\ba
F(\Omega^+_i):  V^*_i\otimes V_i&\rightarrow & {\bf C}, \nonumber \\
x^*\x y&\mapsto& x^*(y), \nonumber \\
F(\Omega^-_i): V_i\otimes V^*_i&\rightarrow &{\bf C}, \nonumber \\
x\x y^*&\mapsto& (-1)^{[x]}y^*(\Kr x) , \nonumber \\
F(U^+_i): {\bf C}&\rightarrow &V_i\otimes V^*_i,\nonumber\\
c&\mapsto& c\sum_t b_t\x b_t^*,  \nonumber\\
F(U^-_i): {\bf C}&\rightarrow &V^*_i\otimes V_i,\nonumber\\
c&\mapsto& c\sum_t (-1)^{[b_t]}b_t^*\x \Kr^{-1} b_t,
\na
where  $\{b_t\}$ is  a basis for $V_i$,
and $\{b_t^*\}$ a basis for $V^*_i$,
which are dual to each other in the sense that  $b_s^*(b_t)=\delta_{st}$.
\end{enumerate}
\end{theorem}

An important property of the functor $F$\cite{ReT} is that
if $\Gamma$ is a coloured ribbon $(k, k)$ graph,
and $\hat\Gamma$ is the coloured ribbon $(0,0)$ graph
obtained by closing $\Gamma$, then
\ban
F(\hat\Gamma)&=&str[\Delta^{(k-1)}(\Kr) F(\Gamma)],
\nan
where the supertrace is taken over the tensor product of the $A$ modules
colouring the open strands of $\Gamma$.  In particular, if $\Gamma$
is a $(1, 1)$ graph with the open strand coloured by such an
$A$ module $V$ that all central elements of $A$ act on it as scalars,
then
\ban
 F(\Gamma)&=& \gamma id_V: V\rightarrow V,\\
F(\hat\Gamma)&=& \gamma SD_q(V),
\nan
where $\gamma$ is a constant.  It follows that for any  $(0,0)$ graph,
if any of its components is coloured with an $A$ module(on which
central elements of $A$ act on as scalars, ), which has a vanishing
$q$ - superdimension,  then the Reshetikhin - Turaev functor yields
zero when applied to it.

Let us now consider as examples the ribbon $(k, \ k)$ graphs depicted
in Figure 2.

\vspace{1 cm}
\begin{center}
Figure  2.
\end{center}
\vspace{1 cm}

\noindent
We colour the bottom ribbons of both Figure $2.a$ and $2.b$ by
$((i_1, +1), (i_2, +1), ..., (i_k, +1))$,
while colour the annulus of Figure 2.a by such an element $j$ of
$I$ that the associated $A$ module $V_j$ has the property that
all central elements of $A$ act on it as scalars.
We denote the resultant coloured ribbon graphs by
$\phi^{(k)}_j$ and $\zeta^{(k)}$ respectively.
It is straightforward to obtain, for $k=1$,
\ban
F(\phi^{(1)}_j)=\chi_j(v^{-1})C_j:
V_{i_1}\rightarrow V_{i_1}, \\
F(\zeta^{(1)})=v:
V_{i_1}\rightarrow V_{i_1}, \\
\nan
where $C_j$ is the central element of $A$ defined by equation (\ref{eq:C})
with the representation afforded by the $A$ module $V_j$.  The
$v\in A$ is defined by (\ref{eq:v}), and $\chi_j(v^{-1})$ is
its eigenvalue in $V_j$.
Either by using induction on $k$ or using the properties of
coloured ribbon graphs discussed in {\em Remarks 6.4.2} of \cite{ReT},
we can show that
\begin{lemma}
\ba
F(\phi^{(k)}_j)=\chi_j(v^{-1})\Delta^{(k-1)}(C_j):
& &V_{i_1}\otimes V_{i_2}\otimes...\otimes V_{i_k}
\rightarrow V_{i_1}\otimes V_{i_2}\otimes...\otimes V_{i_k},
\nonumber\\
F(\zeta^{(k)})=\Delta^{(k-1)}(v):
& &V_{i_1}\otimes V_{i_2}\otimes...\otimes V_{i_k}
\rightarrow V_{i_1}\otimes V_{i_2}\otimes...\otimes V_{i_k}.
\label{eq:kirby}
\na
\end{lemma}
These equations will play a central role in the construction of
three - manifold invariants in the next section.

\section{Three - manifold invariants}
The interelationship between knot theory and the theory of three -
manifolds has long been known. In the early 1960s, Lickorish and
Wallace proved that each framed link in $S^3$ determines
a compact, closed, oriented $3$ - manifold,
and every such $3$ - manifold is obtainable by surgery
along a framed link in $S^3$\cite{Lic}. A framed link
is a link with each of its component associated with an integer.
We will work with the so called blackboard framing throughout
this paper.

However, different framed links may yield homeomorphic $3$ - manifolds
upon surgery. A theorem due to Kirby and Craggs\cite{Kirby} and refined
by Fenn and Rourke in \cite{FR} states that
orientation preserving homeomorphism classes of
compact, closed, oriented $3$ - manifolds correspond bijectively
to equivalence classes of framed links in $S^3$, where
the equivalence relation is generated by the Kirby moves given in
Figure 3.

\vspace{1 cm}
\begin{center}
Figure  3.
\end{center}
\vspace{1 cm}

\noindent
It is also known that the special Kirby $(\pm)$ - moves
$\kappa_\pm^{(0)}$  plus either the Kirby $(+)$ - move $\kappa_+$
or the $(-)$ - move $\kappa_-$ generate the entire
Kirby calculus.  We wish to point out that the Kirby moves are local
operations, namely, each operation only alters a part of a given
framed link while leaving the rest of the link unchanged.

Let $L$ be a framed link(in the blackboard framing) embedded in $S^3$,
which consists of $m$ components $L_i,\ \ i=1,2,...,m$.
Surgery along $L$ gives rise a three - manifold, which we denote by
$M_L$. The basic idea of the Reshetikhin - Turaev construction
of topological invariants of
$M_L$ using quantum groups is to make appropriately weighted sums
of the framed link invariants of $L$ associated with different
representations in such a way that the final combinations are
invariant under the Kirby moves.

The framed link $L$ yields a unique ribbon graph by
extending each of its component $L_i$ to an annulus, which has
$L_i$ itself and an $L_i'$ as its edges, where $L_i'$ is a
parallel copy of $L_i$ such that the link number between the two
is equal to the framing number of the latter. We denote this
ribbon graph by $\Gamma$.

Let $(A, \ R, \ \Kr)$ be a ribbon Hopf superalgebra, and
${\cal V({\bf\Lambda})}= \{V(\lambda) | \lambda\in{\bf \Lambda}\}$
a finite set consisting of such objects of
$\Rep$ that all central elements of $A$ act as scalars in each
$V(\lambda)$, and $SD_q(V(\lam))\ne 0$.
We also require that for each $\lambda\in {\bf \Lambda}$,
there exists a $\lambda^*\in {\bf \Lambda}$ such that $V(\lambda^*)$
is isomorphic to the dual $A$ module $V^*(\lambda)$ of $V(\lambda)$,
and two modules $V(\lambda)$ and $ V(\lambda')$ are isomorphic
if and only if $\lambda=\lambda'$.

Colour the component of $\Gamma$ associated with each $L_i$ by the $A$ module
$V(\lambda^{(i)})$, and let $c=\{\lam^{(1)} , \lam^{(2)}, ..., \lam^{(m)}\}$,
where some $\lambda^{(j)}$'s may be equal.   We denote the resultant
coloured ribbon graph by $\Gamma(L, c)$.
Let  ${\cal C}(L, \Lambda)$ be the set of all distinct $c$'s. We define
\ba
\Sigma(L)&=&\sum_{c\in {\cal C}(L,\Lambda)} \Pi_{i=1}^{m}d_{\lam^{(i)}}
\ F(\Gamma(L,\  c)),
\label{eq:Sigma}
\na
where $d_{\lam^{(i)}}$ are a set of constants satisfying
the following conditions:
\begin{enumerate}
\item $d_\lam =d_{\lam^*}$, $\forall \lam \in {\bf \Lambda}$;
\item Let $C_\lam$ be the central element of $A$ defined by
      (\ref{eq:C}) with $V=V(\lam)$, $\lam\in\Lambda$.
Define
\ba
\delta&=&v-\sum_{\lambda\in\bf{\Lambda}}d_\lambda
\chi_\lambda(v^{-1})C_\lambda.  \label{eq:delta}
\na
Then $\delta$ takes zero eigenvalues in all finite dimensional
irreducible $A$ modules with nonvanishing $q$ -  superdimensions.
\end{enumerate}

$\Sigma(L)$ has the following important properties:
\begin{theorem}\label{main}
$\Sigma(L)$ is independent of the orientation chosen for $L$,
and also invariant under the positive Kirby moves $\kappa_+$
and $\kappa^{(0)}_+$.
\end{theorem}

In order to prove the theorem, we need the following
\begin{lemma}\label{master}
For any finite dimensional $A$ - module $W$,
and a linear $A$ homomorphism $f: W\rightarrow W$,
we have
\ba
Str_W\left(\Kr\delta\circ f\right)&=&0,
\label{eq:trace}
\na
where $\delta$ is as defined by (\ref{eq:delta}),
and $Str_W$ represents the supertrace taken over $W$.
\end{lemma}
{\em Proof}: The $A$ - module $W$ can always be decomposed  into
a direct sum of indecomposable submodules. A direct summand
may contribute to the supertrace only if its image
under $f$ is contained in itself. Also the kernel of $f$
does not contribute to the supertrace, thus we may assume that
$W$ is indecomposable, and the $A$ homomorphism
$f: W\rightarrow W$ is bijective.

If $W$ is irreducible, (\ref{eq:trace}) does not need any proof.
Assume $W$ is not irreducible but admits the composition series
\ban
W\supset W_1 \supset \{0\},
\nan
then we necessarily have $f(W_1)=W_1$, as otherwise $W$ would not
be indecomposable.  We can now rewrite the supertrace
$Str_W\left(\Kr\delta\circ f\right)$
as the sum $Str_{W/W_1}\left(\Kr\delta\circ f'\right)$ $+$
$Str_{W_1}\left(\Kr\delta\circ f|_{W_1}\right)$,
where $f': W/W_1\rightarrow W/W_1$ is the map naturally induced by
$f: W\rightarrow W$, and  $f|_{W_1}: W_1\rightarrow W_1$ is the
restriction of $f$ to $W_1$, which again is bijective.
Since both $W/W_1$ and $W_1$ are irreducible,  the supertraces
in the sum vanish separately.

To proceed further, we use induction on the length of
the composition series for $W$. For a given composition series
\ban
W\supset W_1 \supset W_2\supset ...\supset W_t\supset \{0\},
\nan
we have another equivalent one
\ban
W\supset X_1 \supset X_2\supset ...\supset X_t\supset \{0\},
\nan
where $X_i$ is the image of $W_i$ under $f$. $W_1\cap X_1$ is
again a submodule of $W$, and its image under $f$ is identical
to itself. Let us rewrite the supertrace
$Str_W\left(\Kr\delta\circ f\right)$ as
\ba
Str_W\left(\Kr\delta\circ f\right)&=&
Str_{W/(W_1\cap X_1)}\left(\Kr\delta\circ {\tilde f}\right)
+ Str_{W_1\cap X_1}\left(\Kr\delta\circ f|_{W_1\cap X_1}\right),
\label{eq:composition}
\na
where ${\tilde f}: W/(W_1\cap X_1)\rightarrow W/(W_1\cap X_1)$
is the map induced by $f$, and $f|_{W_1\cap X_1}: W_1\cap X_1$
$\rightarrow W_1\cap X_1$ is the restriction of $f$ to $W_1\cap X_1$.
The second term on the right hand side of (\ref{eq:composition})
vanishes following the induction hypothesis. To consider the first
term, we need to examine the two cases
$W_1=X_1$ and  $W_1\ne X_1$ separately.   In the former case,
$W/(W_1\cap X_1)=W/W_1$ is irreducible, hence
$Str_{W/(W_1\cap X_1)}\left(\Kr\delta\circ {\tilde f}\right)=0$.
In the latter case, $W_1\cap X_1$ is a maximal submodule of both
$W_1$ and $X_1$, and $W=W_1 + X_1$. Also, for any
composition series for $W_1\cap X_1$,
\ban
(W_1\cap X_1)\supset Y_3\supset ... \supset Y_t\supset{0},
\nan
the following
\ban
W\supset W_1\supset(W_1\cap X_1)\supset Y_3\supset ...
\supset Y_t\supset{0},
\nan
is another composition series for $W$.
Now  $W/(W_1\cap X_1)=V^1\oplus V^2$,
where $V^1=W_1/(W_1\cap X_1)$ and $V^2=X_1/(W_1\cap X_1)$.
Clearly, ${\tilde f}(V^1)=V^2$ and ${\tilde f}(V^2)=V^1$.
Hence,  we again have
$Str_{W/(W_1\cap X_1)}\left(\Kr\delta\circ {\tilde f}\right)=0$,
and this completes the proof of the lemma.\\

Now we turn to the  proof of Theorem \ref{main}.
The first statement of the theorem is easy to see:
Reversing the orientation of any component $L_i$ of $L$ is
equivalent to replacing the $A$ module $V(\lam^{(i)})$ colouring $L_i$
by its dual $V({\lam^{(i)}}^*)$. As $d_{\lam^{(i)}}=d_{{\lam^{(i)}}^*}$,
$\Sigma(L)$ is not affected.

To prove the second part of Theorem \ref{main}, we consider the framed
links $L$ and $L'$ given in Figure 4,

\vspace{1 cm}
\begin{center}
Figure 4.
\end{center}
\vspace{1 cm}

\noindent
which are related to each other by a Kirby $(+)$ - move $\kappa_+$.
The $T$ appearing in both $L$ and $L'$ is an arbitrary oriented
$(k, k)$ tangle.  It gives rise to a ribbon graph,
which we colour by $c$ such that the $A$ modules
$V(\lam^{(1)} )$, $V(\lam^{(2)})$, ..., $V(\lam^{(k)})$
are respectively assigned to
the $k$ open strands both on the top and down the bottom.
We denote this coloured ribbon graph by $\Gamma(T;c)$.
The colour of $\Gamma(T;c)$ induces a unique colour for the ribbon
graph associated with $L'$. To colour the ribbon graph
arising from $L$, we also need to assign
$V(\mu), \ \mu\in{\bf\Lambda}$ to the
annulus corresponding to the framing $+1$ unknot.
Now
\ban
F(\Gamma(L; c, \mu))&=&Str_{V(\lam^{(1)})\x...\x V(\lam^{(k)})}
\left\{\Kr F(\phi^{(k)}_\mu)\circ F(\Gamma(T; c))\right\},\\
F(\Gamma(L'; c))&=&Str_{V(\lam^{(1)})\x...\x V(\lam^{(k)})}
\left\{\Kr F(\zeta^{(k)})\circ F(\Gamma(T; c))\right\},
\nan
where we have used equation (\ref{eq:kirby}) and the fact that
$F(\phi^{(k)}_\mu)$ and  $F(\zeta^{(k)})$ both commute with
$F(\Gamma(T; c))$.
Inserting these equations in (\ref{eq:Sigma}) we arrive at
\ban
\Sigma(L)&=&\sum_c \Pi_{i=1}^k d_{\lam^{(i)}}
Str_{V(\lam^{(1)})\x...\x V(\lam^{(k)})}
\left\{\Kr \sum_\mu d_\mu\chi_\mu(v^{-1})C_\mu
\circ F(\Gamma(T; c))\right\},\\
\Sigma(L')&=&\sum_{c} \Pi_{i=1}^{k}d_{\lam^{(i)}}
Str_{V(\lam^{(1)})\x...\x V(\lam^{(k)})}\left\{\Kr v
\circ F(\Gamma(T; c))\right\}.
\nan
Hence
\ba
\Sigma(L') - \Sigma(L)&=&\sum_c \Pi_{i=1}^k d_{\lam^{(i)}}
Str_{V(\lam^{(1)})\x...\x V(\lam^{(k)})}\left\{\Kr\delta\circ
F(\Gamma(T; c))\right\}\nonumber\\
&=&0,
\na
where the last equality follows from Lemma \ref{master}.
The invariance of $\Sigma$ under the special Kirby $(+)$ - move
$\kappa_+^{(0)}$ is a direct consequence of the defining property
of $\delta$.

Let $L$ be an oriented framed link of $m$ components,
and $L'$ be the framed link obtained
by applying once the special Kirby $(-)$ move, namely, adding
a framing $-1$ unknot to $L$.
We consider the matrix $A_L=(a_{ij})_{m\times m}$ defined in the
following way: $a_{ii}$ equals the framing number of the $i$-th
component of $L$, and $a_{ij}$ is equal to the linking number
between the $i$-th and $j$-th component of $L$ if $i\ne j$.
If $M_L$ is the three - manifold obtained
by surgery along $L$, and the $W_L$ is the four manifold bounded
by $M_L$, then the matrix $A_L$ can be interpreted as the intersection
form on the second homology group $H_2(W_L, \ {\bf Z})$.
Let $\sigma(A_L)$ be the number of nonpositive eigenvalues of $A_L$,
then it is clear that $\sigma(A_L)=\sigma(A_{L'})-1$, while the
positive Kirby moves leave $\sigma(A_L)$ unchanged.

Let
\ba
z&=&\sum_{\lam\in{\bf\Lambda}}d_\lam \chi_\lam(v) SD_q(\lam).
\label{eq:z}
\na
If $z\ne  0$,  we define
\ba
{\cal F}(M_L)&=&z^{-\sigma(A_L)}\Sigma(L).
\label{eq:F}
\na

Note that
\ban
\Sigma(L')&=&z\Sigma(L),
\nan
that is, under a special Kirby $(-)$ move, $\Sigma$ is scaled
by $z$.
{}From the properties of $\sigma(A_L)$ discussed above  we conclude
that ${\cal F}(M_L)$  is invariant under the special Kirby $(\pm)$
moves $\kappa_\pm^{(0)}$ and the Kirby $(+)$ move $\kappa_+$.
Since they  generate the entire Kirby calculus,
{\em ${\cal F}(M_L)$ defines a topological invariant of
the three - manifold $M_L$}.

Before closing this section, we remark that any nongraded Hopf algebra
$B$ may be considered as a ${\bf Z}_2$ graded Hopf algebra $A$ with
$A_0=B$, and $A_1=0$. Therefore the construction for three - manifold
invariants developed in this section works equally well in the nongraded
case.  In fact one of the examples which we will study in the
next section is the nongraded quantum group
$U_q(gl(2))$ at odd roots of unity.

The conditions imposed on the $d_\lam$'s are sufficient to ensure
that $\Sigma$ is invariant under the Kirby moves $\kappa_+$ and
$\kappa_+^{(0)}$, as shown by Theorem \ref{main}. To implement
these conditions,  it is only necessary to  study the irreducible
$A$ - modules with nonvanishing $q$ - superdimensions,
or more precisely, the eigenvalues of the central elements $C_\lam$
in these modules.
This of course is still a rather hard problem, as in typical cases,
$A$ is a quantum supergroup at a root of unity, and its irreps are not
well understood\cite{Lusztig}\cite{Concini} in general.
However, for some quantum (super)groups like $U_q(gl(2|1))$ etc.,
the irreps have been completely classified. In these  case,
our construction is easier to apply than that of \cite{RT},
which require, besides other things,  the classification of
indecomposable representations of $A$.

\section{Examples}
In this  section we apply the general method developed in
the last section to  the quantum supergroup  $U_q(gl(2|1))$
at odd roots of unity to explicitly construct topological
invariants of three - manifolds. But for the purpose of
illustrating how the method works, we first consider
$U_q(gl(2))$.

\subsection{Invariants arising from $U_q(gl(2))$ at odd roots of unity}
The three - manifold invariants associated with  $U_q(sl(2))$
at even roots of unity were
constructed by Reshetikhin  and Turaev in Ref. \cite{RT},
and they turned out to be equivalent to Witten's invariants
arising from $su(2)$ Chern - Simons quantum field theory.
These invariants were also explicitly computed for
three - manifolds like the Lens spaces etc. using both
the Chern - Sinoms theory approach\cite{Singer} -- \cite{Rozansky}
and the algebraic approach\cite{Melvin}.  Therefore one has
a reasonably good understanding of these invariants.
Following a similar method as that of Ref. \cite{Lic},
three - manifold invariants were also constructed at odd roots
of unity in \cite{Vogel}. Below is an alternative derivation of
these invraiants. As we will see, some new features also arise
from this exercise.

Let us assume that
\ba
q=exp(i2\pi/N),& N=2r+1, &r\in {\bf Z}_+. \label{eq:q}
\na
The underlying algebra of $U_q(gl(2))$ is generated by
$\{e, f, t_1^{\pm 1}, t_2^{\pm 1}\}$ subject to the constraints
\ba
t_i t_j = t_j t_i, & t_i t_i^{-1} =1, \nonumber \\
t_i e t_i^{-1} = q^{\delta_i,1 -\delta_i,2} e,
&t_i f t_i^{-1} = q^{-\delta_i,1 +\delta_i,2} f,\nonumber\\
{[} e, f] =(k- k^{-1})/(q -q^{-1}),   &k=t_1 t_2^{-1}.
\na
We will also impose the extra relations
\ba
(t_i)^N =1, & (e)^N = (f)^N =0.
\na
The algebra has the structures of a ribbon Hopf algebra with the
co - multiplication
\ban
\Delta(e)&=&e\otimes k+1\otimes e,\\
\Delta(f)&=&f\otimes 1+k^{-1}\otimes f\\
\Delta(t_i^{\pm 1})&=&t_i^{\pm 1}\otimes t_i^{\pm 1};
\nan
the co - unit
\ban
\epsilon(e)= \epsilon(f)=0,& \epsilon(t_i^{\pm 1})=\epsilon(1)=1;
\nan
and the antipode
\ban
S(e)=-e k^{-1},&S(f)=-k f,& S(t_i^{\pm 1})=t_i^{\mp 1}.
\nan
The universal $R$ matrix of $U_q(gl(2))$ reads
\ba
R&=&\sum_{\theta, \sigma, \in {\bf Z}_N}
t_1^{\theta}t_2^{\sigma}\otimes P_1[\theta] P_2[\sigma]
\sum_{\mu\in {\bf Z}_N}
{ {\left[(q-q^{-1})e\otimes f\right]^\mu }\over{[\mu]_q!} },
\na
where
\ban
P_i[\theta]&=&\Pi_{\theta\ne\sigma\in {\bf Z}_N}
             {{t_i-q^\sigma}\over{q^\theta -q^\sigma}}, \\
{[\nu]_q!}&=&\left\{\begin{array}{ll}
         \Pi_{i=1}^\nu[(1-q^{-2i})/(1-q^{-2})], &\nu >0,\\
          1,                                   &\nu=0;
         \end{array}
         \right.
\nan
and
\ban
K_{2\rho}&=&k.
\nan

Each irreducible $U_q(gl(2))$ - module $V(\lam)$ is uniquely
characterized by a highest weight $\lam=(\lam_1, \lam_2)$,
$\lam_1, \ \lam_2$ $\in {\bf Z}_N$. A basis for $V(\lam)$
is given by $\{ v^\lam_i | i=0, 1, ..., n_\lam \}$,
where $ n_\lam\in {\bf Z}_N$ is defined by $n_\lam$
$\equiv \lam_1-\lam_2 (mod N)$.  The actions of the $U_q(gl(2))$
generators on $V(\lam)$ is defined by
\ban
e v^\lam_0 =0, & t_i v^\lam_0=q^{\lam_i} v^\lam_0,\\
f v^\lam_i=v^\lam_{i+1}, & f v^\lam_{n_\lam}=0.
\nan
The $q$ - dimension of  $V(\lam)$ is given by
\ban
D_q(\lam)&=& { {q^{n_\lam +1} - q^{-n_\lam -1}}\over {q -q^{-1}} },
\nan
which vanishes when $n_\lam=N-1$. Therefore, for the purpose of
constructing three - manifold invariants, we only need to consider
the irreps with highest weights in ${\bf\Lam}=\{\lam| n_\lam\ne N-1\}$.

Let $V(\lam^*)$ be the dual $U_q(gl(2))$ module of
$V(\lam)$.  It is easy to see that  $\lam^*=(N-\lam_2, N-\lam_1)$.
Applying $\delta$ to $V(\lam^*)$,
and requiring that its eigenvalue vanish, we arrive at the
following equations
\ban
q^{-\lam_1(\lam_1+1)-\lam_2(\lam_2-1)}&=&
\sum_{\mu\in{\bf\Lam}}d_\mu q^{\mu_1(\mu_1-2\lam_1)+
\mu_2(\mu_2-2\lam_2)} \sum_{\nu=0}^{n_\mu}
q^{2\nu(n_\lam+1)}, \ \ \lam \in {\bf \Lam}.
\nan
Simultaneously replacing $\lam$ by $\lam^*$ and $\mu$ by $\mu^*$
leaves the form of the equations intact, but changing $d_\mu$ to
$d_{\mu^*}$. However, since the equations do not determine
the $d_\mu$'s uniquely, the condition $d_\mu=d_{\mu^*}$ still
needs to be imposed on the solutions.

Solving the equations under the condition $d_\mu=d_{\mu^*}$,
we obtain
\ban
d_\lam &=& q^{ {N^2+1}\over{2} }\left( {G_{N-1}}\over{N}\right)^2
q^{-\lam_1(\lam_1+1)-\lam_2(\lam_2-1)}
\left[q^{\lam_1^2 +\lam_2^2} + x_{n_\lam+1} + x_{N-n_\lam-1}\right],
\nan
where the $x$'s are arbitrary complex numbers, and $G_{N-1}$ is the
$k=N-1$ case of the  Gauss sum defined by
\ban
G_k&=&\sum_{\nu=0}^{N-1} q^{k\nu^2},  \ \ k\in{\bf Z}_N.
\nan
The corresponding $z$ can now be readily worked out, and we have
\ban
z&=& -q^3 \left({{G_{N-1}}\over{\sqrt N}}\right)^4,
\nan
which is independent of the $x$'s.  Inserting the $d$'s and
$z$ in equation (\ref{eq:F}), we arrive at the following
three - manifold invariant
\ban
{\cal F}(M_L)&=&\left\{-q^3\left({{G_{N-1}}\over{\sqrt N}}\right)^4
\right\}^{-\sigma(A_L)}
\sum_{c\in {\cal C}(L, {\bf\Lambda})}
\Pi_{i=1}^{m}d_{\lam^{(i)}}\  F(\Gamma(L,\  c)).
\nan

It is in general a very difficult problem to explicitly compute
this invariant for given manifolds. As a matter of fact,
even the widely studied
Witten -  Reshetikhin - Turaev invariant has only been computed for
some very simple manifolds like the Lens spaces, Seifert manifolds etc..
We hope, in the future, to carry out some explicit computations on this
invariant and the one derived from $U_q(gl(2|1)$ in the next subsection.
Here we merely present a simple example.
For the three -  manifold  obtained by surgery along the framed knot

\vspace{1 cm}
\begin{center}
Figure 5.
\end{center}
\vspace{1 cm}

\noindent
with $k$ crossings and $p$ curls, where $k, p\in {\bf Z}$,
and $k$ is odd, we have
\ban
{\cal F}&=&q^{(N+1)^2/2} G_{2k+2p}
\left({{G_{N-1}}\over{N}}\right)^2
\left\{-q^3\left({{G_{N-1}}\over{\sqrt N}}\right)^4\right\}^{-\sigma(k+p)} \\
&\times&\sum_{\mu=0}^{N-1}q^{(k+p)\mu[1-\mu (N^2-1)/2]-k\mu(\mu+2)-\mu}
\sum_{\nu=0}^\mu (-1)^{\mu+\nu+1} q^{k\nu(\nu+1)}
{ {q^{2\nu+1}-q^{-2\nu-1}}\over{q-q^{-1}} }.
\nan

Note that the free parameters $x_\mu$ miraculously cancel out
among themselves in this particular example.  However, we do not
know whether this is true or not in general. In any case,
the existence of these free parameters is a great advantage:
Appropriately choosing the $x$'s will render some of the  $d$'s
vanishing. Then the corresponding irreps will
not play any role in ${\cal F}$, and can be ignored.
This can greatly simplify the computation of ${\cal F}$.

\subsection{Invariants arising from $U_q(gl(2|1))$}
\subsubsection{$\ur$}
As a  ${\bf Z}_2$ graded algebra over the complex field ${\bf C}$,
the quantum supergroup $U_q(gl(2|1))$ is
generated by $\{e_i, \ f_i, \ t_a, \ t_a^{-1} \ | \ i=1,2, \ a=1,2,3\}$ with
the following relations
\ba
t_a t_b=t_b t_a, & t_a t_a^{-1}=1, & a,  b=1,2,3, \nonumber\\
k_1=t_1 t_2^{-1}, & k_2=t_2 t_3, &k_3=t_1 t_2 t_3, \nonumber\\
k_i e_j k_i^{-1}=q^{a_{ij}} e_j, &
k_i f_j k_i^{-1}=q^{-a_{ij}}f_j, &
{[}e_{i}, f_{j}{\rbrace}=
{ {\delta_{ij}(k_{i}-k_{i}^{-1})}\over{q-q^{-1}} }, \nonumber\\
k_3 e_i k_3^{-1}= e_i, &
k_3 f_i k_3^{-1}= f_i, & i, j = 1,2;     \label{eq:commutation}
\na
\ba
(e_2)^2=(f_2)^2&=&0,\nonumber\\
(e_1)^2 e_2 - ((q-q^{-1})e_1 e_2 e_1 + e_2 (e_1)^2&=&0, \nonumber\\
(f_1)^2 f_2 - (q-q^{-1})f_1 f_2 f_1 + f_2 (f_1)^2&=&0,
\label{eq:serre}
\na
where $q$ is a nonvanishing complex number, and the matrix $(a_{ij})$
is given by
$\left( \begin{array}{rr}
2 & -1 \\
-1 & 0
\end{array} \right)$.
The $[,\ \}$ is the standard graded brackets, and the gradation is
defined by $[t_a]= [e_1]$ $=[f_1]=0$, $[e_2]=[f_2]=1$.

$U_q(gl(2|1))$ has the structures of a
${\bf Z}_2$ graded Hopf algebra, with the
co--multiplication
$\Delta: U_q(gl(2|1)) \rightarrow  U_q(gl(2|1))\otimes  U_q(gl(2|1))$,
\ban
\Delta(e_{i})&=&e_{i}\otimes k_{i}+1\otimes e_{i},\\
\Delta(f_{i})&=&f_{i}\otimes 1+k_{i}^{-1}\otimes f_{i},\\
\Delta(t_a^{\pm 1})&=&t_a^{\pm 1}\otimes t_a^{\pm 1};
\nan
the co - unit
$\epsilon: U_q(gl(2|1)) \rightarrow  \bf C$,
\ban
\epsilon(e_{i})=& \epsilon(f_{i})=&0,\\
\epsilon(t_a^{\pm 1})=&\epsilon(1)=&1;
\nan
and the antipode
$S: U_q(gl(2|1)) \rightarrow  U_q(gl(2|1))$,
\ban
S(e_{i})=-e_{i}k_{i}^{-1},\\
S(f_{i})=-k_{i}f_{i},\\
S(t_a^{\pm 1})=t_a^{\mp 1}.
\nan

In order to construct three - manifold invariants,
we require $q$ be a root of unity. For the sake of simplicity
we again assume that $q$ is  given by (\ref{eq:q}).
Note that although $U_q(gl(2|1))$ is a ribbon Hopf superalgebra
at generic $q$\cite{ZG}, it does not admit a universal $R$ matrix
in the present case.

To circumvent this problem, we observe that the following elements
$\{(e_1)^N,  \ (f_1)^N, $   $(t_a)^{\pm N}-1,\ a=1,2, 3\}$
generate a double sided Hopf ideal $\cal J$ of $U_q(gl(2|1))$,
thus the quotient  $U_q(gl(2|1))/{\cal J}$ is still a ${\bf Z}_2$
graded Hopf algebra, which we denote by $U_q^{(r)}(gl(2|1))$.
The underlying algebra of  $U_q^{(r)}(gl(2|1))$ may be regarded
as defined by (\ref{eq:commutation}), (\ref{eq:serre}) together with
the following relations
\ba
(e_1)^N=(f_1)^N=0, & (t_a)^{\pm N}=1,
\label{extra}
\na
while the co - multiplication $\Delta$, co - unit $\epsilon$ and
the antipode $S$ remain the same.

Now $\ur$ has the structures of a ribbon Hopf superalgebra, with
\ba
\Kr&=&k_2^{-2},  \nonumber\\
R&=&\sum_{\nu, \theta, \sigma, \in {\bf Z}_N}
t_1^\nu t_2^{\theta}t_3^{-\sigma}
\otimes P_1[\nu] P_2[\theta] P_3[\sigma]\   \tilde{R}.
\na
In the above equation
\ban
P_a[\theta]&=&\Pi_{\theta\ne\sigma\in {\bf Z}_N}
             {{t_a-q^\sigma}\over{q^\theta -q^\sigma}}, \\
\tilde{R}&=& \sum_{\theta\in {\bf Z}_N}
{{\left[(q-q^{-1})e_1\otimes f_1\right]^\theta}\over{[\theta]_q!}}
[1\otimes 1 - (q-q^{-1})E\otimes F]
[1\otimes 1 - (q-q^{-1})e_2\otimes f_2],
\nan
where
\ban
E&=&e_1e_2-q^{-1}e_2 e_1, \\
F&=&f_2f_1-q f_1 f_2.
\nan

\subsubsection{Irreducible representations}
In Ref.\cite{Z},  we explicitly constructed all the irreducible
representations of $U_q(gl(2|1))$ using the so - called induced module
construction. From the results of that publication, we can easily extract
all the information about $\ur$ irreps, some main features of which
are listed below:
\begin{enumerate}
\item $\ur$ admits a finite number of irreps, and every irrep
      is finite dimensional;
\item Every irrep is of highest weight type, namely, there exist a
unique highest weight vector and a unique lowest weight vector;
\item The $t_a$'s can be diagonalized simultaneously in every irrep.
\end{enumerate}

To construct an irrep of $\ur$, we start with a one dimensional
module $\{v^\lambda\}$ of the Borel subalgebra generated by
$\{ e_i, t_a\}$ such that
\ba
e_i v^\lambda &=& 0,\ \ \ i=1,2,  \nonumber\\
k_a v^\lambda &=& q^{\lam_a} v^\lambda, \ \ a=1,2,3,
\na
wher $\lam=(\lam_1, \lam_2, \lam_3)$, and we require that
\ban
\lam_a\in {\bf Z}_N, & a=1,2,3.
\nan

Let us now construct an irreducible module $V_0(\lambda)$
over the even quantum subgroup $U_q(gl(2)\oplus u(1))\subset \ur$
generated by $\{e_1, f_1, t_a^{\pm 1}\}$:
\ban
V_0(\lambda)&=&\{(f_1)^j v^\lambda\ |\ i=0,1,...,\lam_1\}.
\nan
Then we build the following $\ur$ module from $V_0(\lambda)$
\ban
\bar{V}(\lambda)&=&V_0(\lambda)\oplus f_2V_0(\lambda)\oplus
F V_0(\lambda)\oplus F f_2 V_0(\lambda).
\nan
Note that $\bar{V}(\lambda)$ is not irreducible in general, but
it is indecomposable, and is generated by a single vector $v^\lambda$.
Therefore all the  central elements of $\ur$ act as scalars in
$\bar{V}(\lambda)$. Let $M(\lambda)$ be the maximal proper submodule
of ${\bar V}(\lambda)$. Define
\ban
V(\lambda)&=& \bar{V}(\lambda)/M(\lambda).
\nan
Then $V(\lambda)$ furnishes an irreducible $\ur$ module.   Let
\ban
Q(\lambda)&=&{{q^{\lam_2}-q^{-\lam_2}}\over{q-q^{-1}}}
{{q^{\lam_1+\lam_2+1}-q^{-\lam_1-\lam_2-1}}\over{q-q^{-1}}}.
\nan
We call $\lambda$ typical if $Q(\lambda)\ne 0$,and atypical otherwise.
Now we give a complete classification of all the irreps.
\begin{enumerate}
\item  For $\lambda$ typical:
\ban
V(\lambda)&=&\bar{V}(\lambda);\\
SD_q(\lambda)&=&0,\\
\chi_\lambda(v)&=&q^{-\lam_3^2-2\lam_2(\lam_1+\lam_2+1)}.
\nan
\item $\lam_2=0$; $v^\lambda$ even:
\ban
V(\lambda)&=& V_0(\lambda)
\oplus \{f_2(f_1)^i v^\lambda | i=1,2,...,\lam_1\},\\
SD_q(\lambda)&=& 1, \\
\chi_\lambda(v)&=&q^{-\lam_3^2}.
\nan
\item $\lam_2\ne 0, \ \lam_1 +\lam_2 +1\equiv 0(mod N)$;
$v^\lambda$ odd:
\ban
V(\lambda)&=& V_0(\lambda) \oplus f_2 V_0(\lambda) \oplus F v^\lambda,\\
SD_q(\lambda)&=&1,\\
\chi_\lambda(v)&=&q^{-\lam_3^2}.
\nan
\end{enumerate}

It is useful to observe that the dual of a typical irrep is also
typical,  and the irreps of dimension greater than one
in the second group are dual to irreps in the third group.

\subsubsection{Three - manifold invariants arising from $\ur$}
Let ${\bf\Lambda}$ be the set of all atypical $\lambda$'s,
namely, ${\bf\Lambda}=\{\lambda=(\lam_1, \lam_2, \lam_3)|
\lam_a \in {\bf Z}_N, \ Q(\lam)=0\}$.
Let $a_{\theta,\omega}$,  $\theta, \omega\in {\bf Z}_N$
be a set of complex numbers satisfying
\ba
a_{0,\omega}&=&a_{0,N-\omega}, \nonumber\\
{{1}\over{N}}q^{\omega^2} G&=&
a_{0,\omega} +\sum_{\theta=1}^{N-1}
(a_{\theta, \omega}+a_{N-\theta, N-\omega}).
\na
Define
\ba
d_{\lambda}&=&q^{-\lam_3^2}
a_{\lam_1,\lam_3}, \ \ \ if \ \ \ \lam_2=0, \nonumber \\
d_{\lambda}&=&q^{-\lam_3^2}
a_{\lam_2, N-\lam_3},
\ \ \ if \ \ \ \lam_1=N-\lam_2-1,\ \ \ \lam_2\ne 0.
\label{eq:d}
\na
Then it can be shown that the central element
\ban
\delta&=&v-\sum_{\lambda\in{\bf \Lambda}}
d_{\lambda} \chi_{\lambda}(v^{-1})C_{\lambda},
\nan
takes zero eigenvalue in all atypical irreps of $\ur$.
It is easy to work out the corresponding $z$ as defined by
(\ref{eq:z}), and we have
\ban
z&=& \left({{G_{N-1}}\over{\sqrt N}}\right)^2.
\nan
Applying (\ref{eq:d}) and $z$ to equation (\ref{eq:F}),
we obtain the following three - manifold  invariant
\ba
{\cal F}(M_L)&=&  \left({{G_{N-1}}\over{\sqrt N}}\right)^{-2\sigma(A_L)}
\sum_{c\in {\cal C}(L, {\bf\Lambda})}
\Pi_{i=1}^{m}d_{\lam^{(i)}}\  F(\Gamma(L,\  c)).
\na

Explicit computations of this invariant for particular  manifolds
of interest is a problem meriting investigation on its own right,
and we hope to return to it in the future.
Here  we consider the special case with $a_{\mu, \nu}=0$,
$\forall \mu\ne 0$.
Now only one - dimensional representations contribute to
the invariant, and this fact makes  ${\cal F}$ very easy to
compute. For any framed link $L$, we have
\ba
{\cal F}(M_L)&=&{{1}\over{N^{m/2}}}
 \left({{G_{N-1}}\over{\sqrt N}}\right)^{m-2\sigma(A_L)}
\sum_{\lam^{(1)}_3,...,\lam^{(m)}_3=0}^{N-1}
q^{\langle\Lam_3|A_L|\Lam_3\rangle},\label{eq:trivial}
\na
where the matrix $A_L$ is as defined in section 4, and
\ban
\langle\Lam_3|A_L|\Lam_3\rangle
&=& \sum_{i, j=1}^m \lam_3^{(i)} (A_L)_{ij}\lam_3^{(j)}.
\nan
Dirct calculations can confirm that (\ref{eq:trivial})
is indeed invariant under all the Kirby moves. It is also worth observing
that this invariant is closely related to the invariant discussed in
\cite{Mattes}, where the latter was also shown to be obtainable
from $U(1)$ Chern - Simons theory.

\vspace{3cm}
\noindent
{\bf Acknowledgements}: I wish to thank Dr. L. G. Kovacs for numerous
helpful discussions on indecomposable representations. This work is
supported by the Australian Research Council.

\pagebreak


\begin{thebibliography}{9999}


\bibitem{Jones} V. F. R. Jones, Bull. Amer. Math. Soc.
    {\bf 12} (1985) 103.
\bibitem{Kauffman} L. H. Kauffman, Topology, {\bf 26}(1987) 395.


\bibitem{Baxter} R. J. Baxter, {\em Exactly solved models in
    statistical mechanics} (Academic Press, 1982).
\bibitem{Yang} C. N. Yang, Phys. Rev. Lett. {\bf 19} (1967) 1312.
\bibitem{Turaev} V. G. Turaev, Invent. Math. {\bf 92} (1988) 527.
\bibitem{Wadati} M. Wadati, T. Deguchi and Y. Akutsu, Phys. Rep.
    {\bf 180} (1989) 248, and references therein.
\bibitem{Witten} E. Witten, Commun. Math. Phys. {\bf 121} (1989) 351.


\bibitem{Drinfeld} V. G. Drinfeld, {\em Quantum groups},
    Proc. ICM, Berkeley, (1986)798.
\bibitem{Jimbo} M. Jimbo, Lett. Math. Phys. {\bf 10}(1985)63.
\bibitem{supergroups} P.P. Kulish and N. Yu Reshetikhin,
    Lett. Math. Phys. {\bf 18} (1989) 143;\\
    M. Chaichian and P.P. Kulish, Phys. Lett. {\bf B234} (1990) 72;\\
    A.J. Bracken, M.D. Gould and R. B. Zhang, Mod. Phys. Lett.
    {\bf A5} (1990) 831; \\
    V. N. Tolstoy, Lect. Notes Phys., {\bf 370} 118(Springer, 1990);\\
    R. Floreanini, V. P. Spiridonov and L. Vinet, Commun. Math. Phys.
    {\bf 137} (1991) 149;\\
    M. Scheunert,  Lett. Math. Phys. {\bf 24} (1992) 173;\\
    R. Floreanini, D. A. Leites and L. Vinet, Lett. Math. Phys.
    {\bf 23} (1991) 127.
\bibitem{ReT} N. Yu Reshetikhin and V. G. Turaev, Commun. Math. Phys.
    {\bf 127} (1990) 1.
\bibitem{Res} N. Yu Reshetikhin, {\em Quantized universal enveloping
    algebras, the Yang--Baxter equation and invariants of links},
    preprints (1987).
\bibitem{ZGB} R. B. Zhang, M.D. Gould and A.J. Bracken,
    {\em Commun. Math. Phys}, {\bf 137} (1991) 13.
\bibitem{Homfly} P. Freyed, D. Yetter, J. Hoste, W.B.R. Lickorish,
    K. Millet and A. Ocneanu, {\em Bull. Am. Math. Soc.} {\bf 12}(1985)239.




\bibitem{Lic} W. B. R. Lickorish, Annals of Math. {\bf 76} (1962) 531.
\bibitem{Kirby} R. Kirby, Invent. Math. {\bf 45} (1978) 35.
\bibitem{FR} R. Fenn and C. Rourke, Topology {\bf 18} (1979) 1.
\bibitem{RT} N. Yu Reshetikhin and V. G. Turaev, Invent. Math.
    {\bf 10} (1991) 547.



\bibitem{Singer} S. Axelrod and I. Singer, {\em Chern - Simons perturbation
    theory}, Proceedings of XXth Conference on Differential Geometrical
    Methods in Physics, World Scientific(1991) $P_{3-45}$.
\bibitem{Freed} D. Freed and R. Gompf, {\em Commun. Math. Phys}
    {\bf 141} (1991) 79.
\bibitem{Jeffrey} L. Jeffry, {\em Commun. Math. Phys}
   {\bf 147} (1992) 563.
\bibitem{Rozansky} L. Rozansky, {\em A large $k$ asymptotics of Witten's
    invariants of Seifert manifolds}, Texas preprint UTTG - 06 - 93.


\bibitem{inv} R. B. Zhang, {Three - manifold invariants arising from
    $U_q(osp(1|2))$}, Mod. Phys. Lett. {\bf A}, in press.
\bibitem{TW} V. Turaev and H. Wenzl, Inter. J. Math. {\bf 4} (1993) 323.
\bibitem{Lic1} W. B. R. Lickorish, Pacific J.  Math. {\bf 149} (1991) 337;
    Math. Ann. {\bf 290} (1991) 657; {\em Calculations with
    the Temperley - Lieb algebra}(preprint).
\bibitem{Viro} V. G. Turaev and O. Y. Viro, Topology {\bf 31} (1992) 865.
\bibitem{Regge} G. Ponzano and T. Regge, in {\em Spectroscopic and group
    theoretical methods in physics}, ed. F. Bloch(North Holland, 1968).
\bibitem{SCS} M. Bourdeau et al, Nucl. Phys. {\bf B372} (1992) 303.
\bibitem{ZG} R. B. Zhang and M.D. Gould, {\em J. Math. Phys.},
    {\bf 32}(1991) 3261.
\bibitem{Z} R. B. Zhang, J. Math. Phys. {\bf 34} (1993) 1236.
\bibitem{Lusztig} G. Lusztig, Geometriae Dedicata {\bf 35}(1990) 89.
\bibitem{Concini} C. De Concini and V. G. Kac, {\em Representations
   of quantum groups at roots of unity}, in {\em Operator algebras,
   unitary representations, enveloping algebras and invariant theory},
   ed. A. Connes et al, Prog. Math {\bf 92} (1990) 471.
\bibitem{Melvin} R. C. Kirby and P. Melvin, Inv. Math.
    {\bf 105} (1991) 473.
\bibitem{Vogel} C. Blanchet, N. Habegger, G. Masbaum and P. Vogel,
    Topology, {\bf 31} (1992) 685.
\bibitem{Mattes} J. Mattes, M. Polyak and N. Reshetikhin,
    {\em On invariants of $3$ - manifolds derived from Abelian groups},
    preprint.

\end{thebibliography}
\end{document}